\documentstyle[11pt,newpasp,twoside,psfig]{article}
\makeindex
\index{a:Weilbacher, P.}
\index{a:Fritze--v. Alvensleben, U.}
\index{a:Duc, P.--A.}
\index{k:Tidal Dwarf Galaxies!census}
\index{k:Tidal Dwarf Galaxies!stars}
\index{k:Tidal Dwarf Galaxies!kinematics}          
\index{k:Tidal Dwarf Galaxies!metallicity}

\def\edcomment#1{\iffalse\marginpar{\raggedright\sl#1\/}\else\relax\fi}
\reversemarginpar

\begin{document}
\title{Stellar Populations of a Sample of Tidal Dwarf Galaxies}
\author{Peter M.~Weilbacher}
\affil{Department of Physics,
       Durham University, 
       South Road, 
       Durham, 
       DH1\,3LE, 
       UK}
\author{Uta Fritze--v.\,Alvensleben}
\affil{Universit\"ats-Sternwarte G\"ottingen,
       Geismarlandstr.~11,
       D-37083 G\"ottingen,
       Germany}
\author{Pierre--Alain Duc}
\affil{CNRS FRE 2591 and CEA, DSM, DAPNIA,
       Service d'Astrophysique,
       Centre d'Etudes de Saclay,
       91191 Gif-sur-Yvette Cedex,
       France}

\begin{abstract}
We investigate the stellar populations of a sample of Tidal Dwarf
Galaxies, combining observations and evolutionary synthesis models to
try and reveal their formation mechanism.
On optical images we select a first sample of TDGs for which optical
spectroscopy is used to measure metallicities and velocity structure.
Finally, we estimate ages, burst strengths, and stellar masses from
near-infrared imaging in comparison with a dedicated grid of evolutionary
synthesis models, to assess if Tidal Dwarfs are formed out of collapsing
gas clouds or by an accumulation of old stars from the parent galaxy or
by a combination of both.
\end{abstract}

\section{Introduction \& Motivation}
The old idea of Zwicky(1956) that (dwarf) galaxies could be formed
during encounters of giant galaxies has received a lot of observational
support within the last decade (Duc \& Mirabel, 1994, Duc \& Mirabel, 1998, Hibbard et~al., 1994). 
This data revealed the so called {\it Tidal Dwarf Galaxies} ({\bf TDG}s)
to be gas rich, dwarf galaxy sized knots with stellar masses of
$10^6\dots10^8$\,M$_\odot$ made of recycled material, still embedded in
the tidal tail of their parent galaxy. Often, a young starburst on top
of an older stellar population is observed. This is widely interpreted
as formation of a new galaxy ``in situ'' in the tidal tails.

To tell the phenomenon of TDGs apart from e.g.~super star clusters
or chance alignments of matter, the following definition of TDGs was
introduced (Duc et~al., 2000; Weilbacher \& Duc, 2001):\\
A Tidal Dwarf Galaxy is {\it a self-gravitating object with the mass of
a dwarf galaxy, formed from recycled material that was expelled from a
big galaxy in the course of tidal interaction}.

Why are TDGs thought to be significant? First, they represent a special
mode of galaxy formation which can be observed in the local universe,
i.e.~much more detailed than high-redshift galaxies forming in the early
universe. Second, part of the dwarf galaxy populations of the Local Group
or nearby galaxy clusters may have been formed as TDGs 
(Okazaki \& Taniguchi, 2000). And
finally, they may explain part of the faint blue galaxy population at
intermediate redshifts. At those times, the merger rate was higher than
today (Conselice et al., 2003) and the galaxies were more gas rich, so that a
large number of bright TDGs could have been formed.

But how do TDGs form? Are they preferentially built as stellar knots of
which there may be several per merger as seen in the dynamical models
of Barnes \& Hernquist (1992), or from giant gas clouds at the tips of the tidal tails
which then afterwards may also attract stars into their potential as in
the simulations of Elmegreen et~al. (1993)? This question was the original starting
point of our project. We therefore investigate the stellar populations of
a sample of TDGs in order to find out which of these models is realized
or more common in nature.

\section{Sample \& Approach}
Our sample of interacting galaxies consists of 14 interacting/merging
galaxies from the catalog of Arp \& Madore (1987). It includes interacting
galaxies of all Hubble types from ellipticals to irregular galaxies,
some of which already in an advanced stage of merging. They have distances
of $\sim50$ to $370$\,Mpc\footnote{converted from their heliocentric
velocities using $H_0=75$\,km\,s$^{-1}$\,Mpc$^{-1}$} and were selected
on the basis of their disturbed appearance. They do not, however, show 
especially
long tidal tails or prominent knots on images of the Digital Sky Survey,
as some of the other galaxies which had TDGs investigated previously.

Our analysis consists of three observational stages, each step is analyzed
by means of evolutionary synthesis models: optical imaging for sample
selection of TDG candidates, optical spectroscopy of these candidate
TDGs to check their physical association with the interacting system
and to investigate the properties and kinematics of the ionized gas,
and, finally, near-infrared imaging to determine burst strengths, ages,
and stellar masses of the TDGs. The observations were carried out with
the ESO NTT, the ESO-3.6m telescope, and the FORS2 instrument on the VLT.

We use a version of the G\"ottingen Galaxy Evolution code GALEV for
our modeling (for the details of the models see Weilbacher et al., 2000; Weilbacher, 2002).
Here, we just recall that we model a starburst of varying strength on top
of an old stellar population, and compute the optical to near-infrared
broad-band luminosities and spectra of these composite populations at
different ages before and after the burst maximum. We note that our
code also includes the effects of gaseous emission lines and continuum
on the spectra and broad band colors, which is an important ingredient
as it can contribute up to 70\% of the luminosity in optical broad band
filters in the strong starbursts that we investigate here. The burst
strength, which we define as the mass ratio \begin{displaymath} b =
\frac{M({\rm young\ stars})}{M({\rm all\ stars})} \end{displaymath}
is a central property in our investigation. Our grid of models covers
burst strengths in the range from 8\% to 100\% of young stars.

\section{Results}
In the following we describe in more detail the procedure and results of
our three stage approach to select and characterize Tidal Dwarfs.

\subsection{Optical Imaging \& Models}
Our optical $B$,$V\!$,$R$ imaging data was analyzed with aperture
photometry yielding color and (absolute) magnitude information on
more than 100 knots in the tidal tails of our sample of interacting
galaxies. To select the best candidates for TDGs we compared this data
with our models in two-color diagrams ($B-V$~vs.~$V-R$). Many objects
turned out to be background objects on the basis to their colors,
several knots were excluded as TDG candidates because of their location
too close to the parent nuclei. We selected {\bf 44 good candidates for
TDGs} with luminosities $-17.0 < M_B < -10.5$\,mag.

At this stage we had to make assumptions about metallicity and
extinction in the TDG candidates, so we could only derive first order
estimates of burst ages and strengths.
Observed color could only be matched by models with fairly high burst
strengths, $b$ up to 20\%. We therefore expect strong fading after the
TDGs exhaust their gas supplies and terminate their current starbursts
(Weilbacher et~al., 2000).

\subsection{Optical Spectroscopy}
The main motivation for our spectroscopic campaign was to confirm the
association of the selected TDG candidates with the main interacting
galaxy and to determine their metallicities and internal extinction
values to further refine our models, and to exclude low metallicity dwarf
galaxies in the neighbourhood of the interacting galaxies. We therefore
chose to observe the whole optical wavelength range with low resolution,
including all important emission lines from [O{\sc ii}]3727 to [S{\sc
ii}]6717,6731 (Weilbacher et~al., 2003).

While we could not get redshifts for all TDG candidates due to constraints
of the slit placements in our multi-object spectroscopy, and in very
few cases also because of low S/N of the resulting spectra, we could
confirm the association of {\bf 29 TDG candidates} with their parent
galaxies. This proves our method with color selection by evolutionary
models as a very successful pre-selection for TDG candidates. {\it All}
of the TDG candidates for which we actually obtained a spectroscopic
redshift are related to the central galaxy!

The metallicity measurement was carried out with various methods used
in the literature for optical emission line spectra, ranging from
the $P$/$P_3$ method of Pilyugin (2001), to the ``standard'' $R_{23}$
method, and the [N{\sc ii}] method of van Zee et~al.(1998). Only in one case
could we use the physical method (Shields, 1990), because the weak line
[O{\sc III}]4363 was undetected in all other cases. We compared the
results of these methods, and if they agreed within their error bars we
selected the value given by the method with the smallest systematical
error. This way, we found three galaxies with $12+\log(O/H) < 8.0$
which we excluded as TDG candidates because ``recycled'' gas cannot allow
such low metallicity TDGs to form. We are therefore left with {\bf 26 TDG
candidates}. These have a mean oxygen abundance in the range $12+\log(O/H)
= 8.34\pm0.14$ or about 1/4\,$Z_\odot$, absolute magnitudes of $M_B =
-12\dots-17$\,mag, and H$\alpha$ luminosities between those of the most
luminous H{\sc ii} regions of spiral galaxies and those of ``normal''
dwarf galaxies.

Finally, and a bit to our surprise, we could also derive velocity curves
of the ionized gas within some of our TDG candidates even at the low
spectral resolution used ($R < 400$). Velocity differences
$\Delta V_{\rm max}$ we measure within these TDGs can reach up to $\sim
500$\,km\,s$^{-1}$. While we do not think that it would be correct to
convert the measured values of $\Delta V_{\rm max}$ into dynamical mass
estimates, and the origin of these apparent velocity gradients is not
clear, it seems reasonable to assume that these knots are kinematically
decoupled from the tidal tail, i.e.~{\it real TDGs}.

\index{o:AM 1353--272}
One of the interacting galaxies from our sample, AM\,1353-272 which we
dubbed ``The Dentist's Chair'' due to its appearance on optical images,
was reobserved with FORS2 on the ESO VLT at higher spectral resolution
($R\approx700$). Curved slits allowed us to cover both tidal tails of
this system and we indeed observed significant velocity gradients in 7
TDGs in this one interacting system (Weilbacher et~al., 2002)
 and $\Delta V_{\rm
max}$ in the range 24 to 343\,km\,s$^{-1}$! This gives us a total number
of {\bf 13 TDGs} with hints of kinematical independence, for now a
lower limit which has to be checked with more spectroscopic data and
higher resolution.

\subsection{Near-Infrared Photometry \& Refined Models}
Our near-infrared (NIR) photometry is mainly derived from deep $H$-band
images taken with the SOFI camera on the ESO NTT. Two of our interacting
systems were also observed in $J$, and three in the $K\!s$-band.
NIR-observations are particularly important to constrain the contribution
of old stars, inherited from the parent galaxy, as these are hidden in
the current starburst at optical wavelengths while they possibly dominate
the NIR luminosities.

This data together with our previous observations gives us the spectral
energy distribution of all our TDGs and TDG candidates in the wavelength
range from $\sim$0.43 to $2.16\,\mu$m.  From spectroscopy we have
another observable, the H$\beta$ equivalent width, so
that we can altogether compare up to 7 observed properties
($B$,$V\!$,$R$,$J$,$H$,$K_{\rm s}$,H$\beta$) with our grid of evolutionary
synthesis models.


To compare the data to the models we developed a least-squares based
procedure, to identify the model that best represents the observed data
in an objectively reproducible way. First, we deredden the observed
colors using the spectroscopic estimate of the total extinction. Next,
we select the model(s) with the metallicity matching the observed oxygen
abundance. Then, we select the timesteps of each model where EW(H$\beta$)
matches the observed value within the error bars, and evaluate how close
the model is to the observed SED for each timestep of these selected
models. Finally, we derive the interesting properties from the model
and timestep best matching the observed properties.

This procedure confirms the rough estimates we previously had from
the comparison with the optical data: our TDGs show young burst ages,
mostly up to 20\,Myr, some even seem to be at the beginning of their
starburst, i.e.~before reaching the maximum star formation rate. The
model also allows us to derive a good estimate of the total stellar masses
which have a mean value of $\sim 2 \times 10^8\,M_\odot$, typical for dwarf
galaxies. We also estimate the current star formation rate (SFR) in the
TDGs, circumventing the problems with the usual $L($H$\alpha)$ conversion
(Weilbacher \& Fritze-von Alvensleben, 2001). The mean SFR of 
0.05\,$M_\odot\,$yr$^{-1}$ shows them to
be strongly star-forming galaxies in relation to their sizes and masses.

\begin{figure}[t]
\centerline{\psfig{file=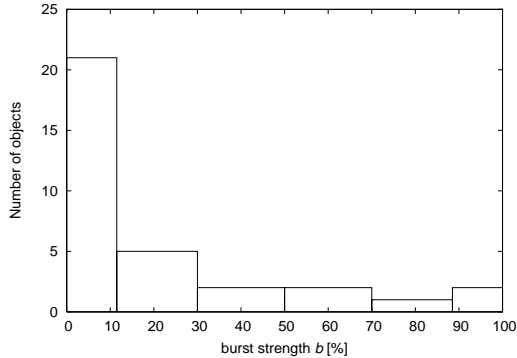,width=7cm}}
\caption{A histogram of the current burst strength derived from the
         comparison of evolutionary models with the data of our sample
         of TDGs.}\label{b-TDGs}
\end{figure}

Most importantly, we also derive the burst strengths $b$:
Fig.~\ref{b-TDGs} show the distrubtion of burst strenghts of the TDG
candidates in our sample. Their current burst strength $b$ indeed covers
the entire range of possible values from 0 to 100\%, with a strong peak
in the lowest bin ($b<10\%$). This shows that many TDGs in our sample
have most likely formed from stellar condensations and only few of them
from the collapse of giant gas clouds (Weilbacher et al.,~in prep.).


\section{Conclusions}
We used a three stage approach of optical photometry, optical
spectroscopy, and near-infrared photometry combined with dedicated
evolutionary synthesis models to select a sample of TDG (candidates),
and study their gaseous properties and stellar populations.

We selected 44 TDG candidates from optical imaging in comparison with
our models from over 100 knots in the tidal tails of our sample of 14
interacting and merging galaxies. From optical spectroscopy we confirm
the association of the TDG candidates with the parent galaxies. The
metallicity measurement shows them to have a mean oxygen abundance close
to 1/4\,$Z_\odot$, while three TDG candidates were rejected due to their
too low metallicities.
Low resolution spectroscopy also revealed 13 TDGs to show kinematical
signatures independent of their surrounding tidal tail. We therefore
roughly find at least one true TDG per interacting system.

From our near-infrared analysis and the detailed comparison of their
spectral energy distributions with our evolutionary synthesis models
we confirm that the TDGs in our sample indeed have the stellar masses
of dwarf galaxies and show strong star-formation activity. Their burst
strengths show most of them to have primarily formed from stellar
condensations and only a smaller percentage from the collapse of giant
gas clouds.

\acknowledgments{PMW's work on this project was partially supported by the DFG
(grants FR 916/6-1 and 916/6-2). Further support by the European Research
and Training Network {\it Spectroscopic and Imaging Surveys for Cosmology}
under contract HPRN-CT-2002-00316 is acknowledged. UFvA gratefully
acknowledges partial travel support from the IAU \& DFG (Fr 916/12-1).}


\end{document}